\documentclass[epsf,floatfix,prl,aps,twocolumn]{revtex4}

\usepackage{epsf}

\begin{document}

\preprint{DRAFT-NOT FOR CIRCULATION}

\date{May 29, 2003}

\title{Continuum description of profile scaling in nanostructure decay}


\author{Dionisios Margetis$^1$, Michael J.~Aziz$^2$, and Howard A.~Stone$^2$}

\affiliation{${}^1$Department of Mathematics, Massachusetts Institute of Technology, Cambridge, MA 02139\\
${}^2$Division of Engineering and Applied Sciences, Harvard University, Cambridge, MA 02138}


\begin{abstract} The relaxation of axisymmetric crystal surfaces with a single facet
                 below the roughening transition is studied via a continuum
                 approach that accounts for step energy $g_1$ and
                 step-step interaction energy $g_3>0$. For diffusion-limited
                 kinetics, free-boundary and boundary-layer
                 theories are used for self-similar shapes close to the growing facet.
                 For long times and $\frac{g_3}{g_1} < 1$, (a) a universal equation is derived for the shape
                 profile, (b) the layer thickness varies as $(\frac{g_3}{g_1})^{1/3}$, (c) distinct solutions
                 are found for different $\frac{g_3}{g_1}$, and (d) for conical shapes,
                 the profile peak scales as $(\frac{g_3}{g_1})^{-1/6}$.
                 These results compare favorably with kinetic simulations.
\end{abstract}

\maketitle

\section*{}

The drive toward smaller features in devices has fueled much
interest in low-temperature kinetic processes such as growth,
etching, and morphological relaxation.  The constantly decreasing
temperatures present increasing challenges for treatment of
thermodynamics, kinetics and macroscopic evolution of surfaces. A
crystal surface at equilibrium undergoes a roughening transition
at a surface orientation-dependent temperature
$T_R$~\cite{burtoncabrera,wortis}. In equilibrium at temperature
$T$, crystal facets (planar regions of the surface) have $T_R >
T$, whereas orientations in continuously curved portions of the
surface have $T_R < T$.  In numerous nonequilibrium situations
below $T_R$, a crystal surface relaxes to its equilibrium shape
via the lateral motion of steps at a rate limited mainly by the
diffusion of adatoms across terraces and attachment and detachment
at step edges. Here we report a continuum treatment of this
evolution using a partial differential equation (PDE) and obtain
scaling laws and universal aspects of the solutions.

Morphological equilibration for surfaces above $T_R$
is described by a continuum treatment~\cite{mullins57,mullins59}
where the surface free energy, which is an analytic function of
the surface slope, and chemical potential~\cite{herring51a}
are ingredients in a fourth-order PDE for the evolution of
the surface profile. However,
this analysis is not applicable to surfaces below $T_R$ because
the surface free energy is not analytic in the
surface orientation~\cite{herring51b,grubermullins,wortis}; see
Eq.\,(\ref{G-eqn}) below.

Efforts to describe morphological evolution below $T_R$ began in
the mid 1980s and include simulations of the motion of monatomic
crystalline steps and continuum thermodynamic approaches.  In the
latter, to account for evolution due to the motion of steps
separating terraces below the basal plane's $T_R$, the step
density, which is proportional to the surface slope, is introduced
as a variable within a coarse-grained continuum description on a
scale large compared to the step separation (typically, 1-10 nm).
Expressions have been developed for the chemical potential of
atoms at interacting step edges, leading to a nonlinear PDE for
evolution of periodic surface
modulations~\cite{rettorivillain,lanconvillain}; progress has been
made toward solving the PDE~\cite{ozdemirzangwill90,jeongwilliams}
but has been hindered when evolution involves facets~\cite{spohn93,shenoyfreund}.
Kinetic simulations mimic nanoscale processes and so have been used to describe the
detailed motion of many steps, as well as the evolution of facets, via coupled differential
equations~\cite{israelikandel99,israeli00,israelikandel00}.
Nevertheless, the kinetic simulations are generally limited in their
ability to characterize universal features of the shape evolution.
Here we show that the shape profile, including the facet, can be
treated using a continuum, thermodynamic description that
illuminates scaling aspects of the kinetic behavior; for this purpose, we use
an analytical framework that transcends the limitations of
continuum approaches previously recognized
~\cite{israelikandel02}.

Our analytical approach treats facet evolution as a free-boundary
problem~\cite{spohn93,hagerspohn}. The surface height is $h({\bf
r}, t)$, where ${\bf r}=(x,y)= r {\bf e}_r$ is the position vector
in the plane of reference; see Fig.~1 for an axisymmetric shape.
The step density is $|\nabla
h|; \nabla h \equiv (h_x,h_y)$ and subscripts denote partial
derivatives. Denoting the atomic volume by $\Omega$ and the
surface current (atoms per length per time) by ${\bf j}$, the
conservation equation for adatoms is
\begin{equation}
\frac{\partial h}{\partial
t}+\Omega\nabla\cdot {\bf j} = 0 ~.  \label{continuity}
\end{equation}
The current is ${\bf j} = - c_s D_s \nabla \mu/k_B T$, where $D_s$
and $c_s$ are the surface diffusivity and adatom
concentration, and $\mu({\bf r})$ is the chemical potential of
atoms at step edges; $\,D_s$ is in principle a tensor function of $\nabla h$.

\begin{figure}
\centerline{\epsfxsize0.35\textwidth\epsfbox{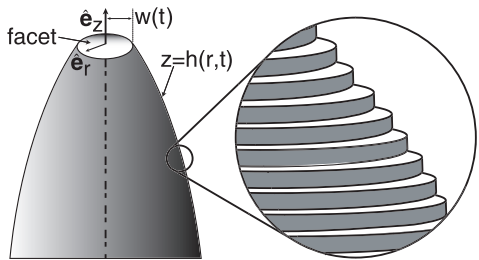}}
\caption{View of an axisymmetric surface profile, on both
the macroscale and the nanoscale where the atomic steps
are evident. The evolution of surface morphology is
caused by the motion of steps.}
\end{figure}

We focus on diffusion-limited (DL) kinetics, where diffusion of
adatoms across terraces is the rate-limiting process,
and further assume that $c_s$ is
constant and $D_s$ is a scalar constant. Equation (\ref{continuity}) becomes
\begin{equation}
\frac{\partial h}{\partial
t} = \frac{c_s D_s\Omega}{k_B T} \nabla^2 \mu ~.  \label{continuity1}
\end{equation}
Next, $\,\mu$ and $h$ are related via the surface free energy per
unit projected area, $\,G$. A common expression for $G$ of
vicinal surfaces for $T< T_R$ assumes that $G$ depends on the step
density according
to~\cite{jeongwilliams,grubermullins,jayaprakashetal}
\begin{equation}
G(\nabla h) = g_0 + g_1 |\nabla h| + \frac{1}{3}g_3
|\nabla h|^3~.  \label{G-eqn}
\end{equation}
The $g_0$ term represents the surface free energy of the reference
plane, $\,g_1$ is the step energy (line tension), and $g_3$, which
accounts for step-step interactions, includes entropic repulsions
due to fluctuations at the step edges and pairwise energetic
interactions between adjacent steps. All $\,g_0$,$\,g_1$ and $g_3$
are temperature dependent and we consider repulsive interactions
between steps, $\,g_3>0$.

The surface chemical potential is derived from $G$ by the
relation~\cite{herring51a,grubermullins,ozdemirzangwill90}
$\mu =-\Omega\nabla \cdot \frac{\delta G}{\delta (\nabla h)}$,
where $\frac{\delta}{\delta(\nabla h)}\equiv\left(\frac{\partial}{\partial h_x},\,\frac{\partial}{\partial
h_y}\right)$~\cite{funcderivative}:
\begin{equation}
\mu =-\Omega g_1 \,\nabla\cdot\left [ \left ( \frac{\nabla
h}{|\nabla h|}\right ) +\frac{g_3}{g_1} \left (|\nabla h|\nabla h\right )
\right ]~. \label{mu-eqn}
\end{equation}

The surface evolution equation follows by combination of Eqs.\
(\ref{continuity})--(\ref{mu-eqn}). We use cylindrical coordinates
to describe the relaxation of axisymmetric shapes $h=h(r,t)$
that are smooth along the surface outside the facet and have
negative slope,$\,\frac{\partial h}{\partial r}<0$ (Fig.~1).
Since $\nabla h = {\bf e}_r \frac{\partial h}{\partial r}$,
it is convenient to define the dimensionless step density or surface slope
$F(r,t) = - \frac{\partial h}{\partial r}$. The surface
then evolves according to a fourth-order nonlinear PDE for $F$,
\begin{equation}
\frac{\partial F}{\partial t} = \frac{3B}{r^4}-
B\,\frac{g_3}{g_1}\, \frac{\partial }{\partial r}\,
\nabla^2 \left [\frac{1}{r} \frac{\partial }{\partial r}
\left (r F^2 \right ) \right]~.
\label{governing-eqn}
\end{equation}The material parameter
$B=\frac{c_s D_s \Omega^2 g_1}{k_B T}$ has dimensions (length)$^4$/time.

Equation (\ref{governing-eqn}) is supplemented with the initial
condition $F(r,0)=-H^{\prime}(r)$, where $H(r)=h(r,0)$ is the
initial surface profile with the properties $H^{\prime}(r)=0$ for
$r<W$ (the initial facet radius) and $H'(r)<0$ for $r>W$. Also, there
are four boundary conditions applied at the facet edge,$\,r=w(t)$.  In particular,
the height $h$ and the current
${\bf j}$ are continuous at the facet edge.  The
latter condition, along with $r{\bf j}\to {\bf 0}$ as
$r\to\infty$, ensures that the total mass is conserved. A
consequence of Eq.\ (\ref{governing-eqn}) and the initial
conditions is that no other facets are formed.  Another condition
is slope continuity at the facet edge,$\,F(w,t)=0$ (i.e., local
equilibrium~\cite{jayaprakash83,bonzel01} and it is also
consistent with kinetic simulations~\cite{israelikandel99}).
It is shown below that Eq.\ (\ref{governing-eqn}) furnishes
$F(r,t)=O(\sqrt{r-w})$ as $r\to w^+$ where the coefficient is time
dependent~\cite{phase}. Finally, for $r\le w$, where there is a
facet, we extend continuously through
Eqs.~(\ref{continuity})-(\ref{mu-eqn}) the
variable $\mu$, although it no longer represents the true chemical
potential, and the quantity ${\bf N}={\bf e}_r N\equiv
-\frac{\delta G}{\delta(\nabla h)}$ whose divergence yields
$\mu/\Omega$~\cite{spohn93,margetisetal}. Hence the final two
boundary conditions are continuity of $\mu$ and of ${\bf N}$. Although we
now have a mathematically complete set of boundary conditions for
Eq.\ (\ref{governing-eqn}), the issue of the boundary conditions
remains a topic of discussion~\cite{chame96,shenoyfreund02}. Nevertheless, we expect
that the main analytical and scaling ideas given below are
independent of the detailed form of these conditions.

The boundary conditions described above relate $\frac{g_3}{g_1}$ to the derivatives
of $F^2$ at $r=w^+$, the facet height $h_f(t)$
and the facet radius $w(t)$.
By differentiating $h_f(t)=h(w^+(t),t)$ in time, we deduce that, at $r=w$,
\begin{equation}
B\{1-\frac{g_3}{g_1} w[(F^2)'-2w (F^2)''-w^2 (F^2)''']|\}=\dot h_f w^3,\label{cond1}
\end{equation}
where the dot denotes the time derivative.
Then, an examination of $\mu$ and $\bf N$ and their continuous extensions on the
facet gives two more conditions at $r=w$~\cite{margetisetal}:
\begin{equation}
w[3(F^2)'-w^2(F^2)''']=3\frac{g_1}{g_3}=w[3(F^2)'-w(F^2)''].\label{conds}
\end{equation}

We now treat Eq.\ (\ref{governing-eqn}) with conditions
(\ref{cond1}) and (\ref{conds}) and $F=0$ at the facet edge as a
free-boundary problem~\cite{spohn93}: there is a moving facet for
$r<w(t)$, where $F=0$, and this facet connects smoothly to the
rest of the profile for $r> w(t)$. Note that this problem
statement is valid for arbitrary $\frac{g_3}{g_1}$. In general,
there exist an ``outer" region, where only the line-tension energy
$g_1$ is important, and an ``inner" region in the neighborhood of the facet edge,
where the step-step interaction energy $g_3$ becomes significant.
Motivated by kinetic simulations with
$\frac{g_3}{g_1}<1$~\cite{israelikandel99}, we set
$\epsilon\equiv\frac{g_3}{g_1}$ and treat Eq.\
(\ref{governing-eqn}) analytically for the case with small
$\epsilon$, i.e., weak repulsive interactions. Because the small
parameter $\epsilon$ multiplies the highest-order spatial
derivative in Eq.\ (\ref{governing-eqn}), the shape evolution can
be treated with boundary-layer theory~\cite{hinch}.  We
start with the solution for $\epsilon=0$ where the corresponding
facet radius $w(t; \epsilon)$ is denoted $w(t;0)=w_0(t)$. From Eq.\
(\ref{governing-eqn}), the zeroth-order outer solution $F(r,t;
0)\equiv F_0 (r,t)$ satisfies ${\partial F_0}/{\partial t} =
3B/r^4$, which is integrated subject to the initial condition
$F_0(r,0) = -H^\prime (r)$ to give
\begin{equation}
F_0 (r,t) = 3Bt\,r^{-4}
-H^\prime (r) ~,\qquad r> w_0(t) ~.\label{F0-eqn}
\end{equation}
At the facet edge, $\,F_0(w_0,t) = 3Bt/w_0^4 -H^\prime (w_0)\ne 0$, so the
slope profile is discontinuous; this feature motivates a singular perturbation
analysis.

The next step is to examine how the inclusion of a nonzero
$\epsilon$ renders the slope continuous by retaining the highest
derivative in Eq.\ (\ref{governing-eqn}). We therefore consider a
region of width $\delta (t;\epsilon)\ll w$ in the neighborhood of
the moving facet edge, and describe the solution in this region in
terms of the local variable $\eta \equiv (r-w)/\delta$. Thus, we
seek a long-time similarity solution that depends on the distance
from the facet edge and time, $\,F(r,t;\epsilon) = {\cal F}(\eta,
t;\epsilon)$. We anticipate that, to leading order in $\epsilon$,
\begin{equation}
{\cal F} (\eta, t)\sim  a_0(t) f_0(\eta; \epsilon)~,\qquad
\eta=[r-w(t;\epsilon)]/\delta(t;\epsilon)~, \label{similarity-soln}
\end{equation}
where $f_0$ depends implicitly on $\epsilon$ through the boundary conditions.
Substitution of (\ref{similarity-soln}) into (\ref{governing-eqn})
and balance of the leading-order terms in $\epsilon$ gives
\begin{equation}
\frac{{\dot w} \delta^3}{B\epsilon a_0} f_0^{\prime} =
(f_0^2)^{\prime\prime\prime\prime}
+ O\left (\frac{\delta}{w}, \frac{\delta^4}{\epsilon w^4},
\frac{{\dot \delta} \delta^3}{\epsilon B} \right )~.\label{f0-eqn1}
\end{equation}
Thus,  $\,\frac{\dot w \delta^3}{B\epsilon a_0}$ must be
time-independent and we take it equal to unity without affecting
observable quantities such as $F$ or $w$. It follows that
$\delta(t;\epsilon) = O(\epsilon^{1/3})$, independent of the
(axisymmetric) initial conditions, which is a prediction for a
scaling law for the boundary-layer width in the case of DL
kinetics. The neglected terms in Eq.\ (\ref{f0-eqn1}) are
$O(\epsilon^{1/3})\ll 1$. Furthermore, the leading-order facet
radius is $w(t)^4\sim 4B\int_0^t dt'\,\tilde w(t')^3 a_0(t)$,
where $\tilde w=\frac{w}{\Delta}$ and
$\Delta(t)\equiv \epsilon^{-1/3}\,\delta(t;\epsilon)$.

\begin{figure}
\centerline{\epsfxsize0.35\textwidth\epsfbox{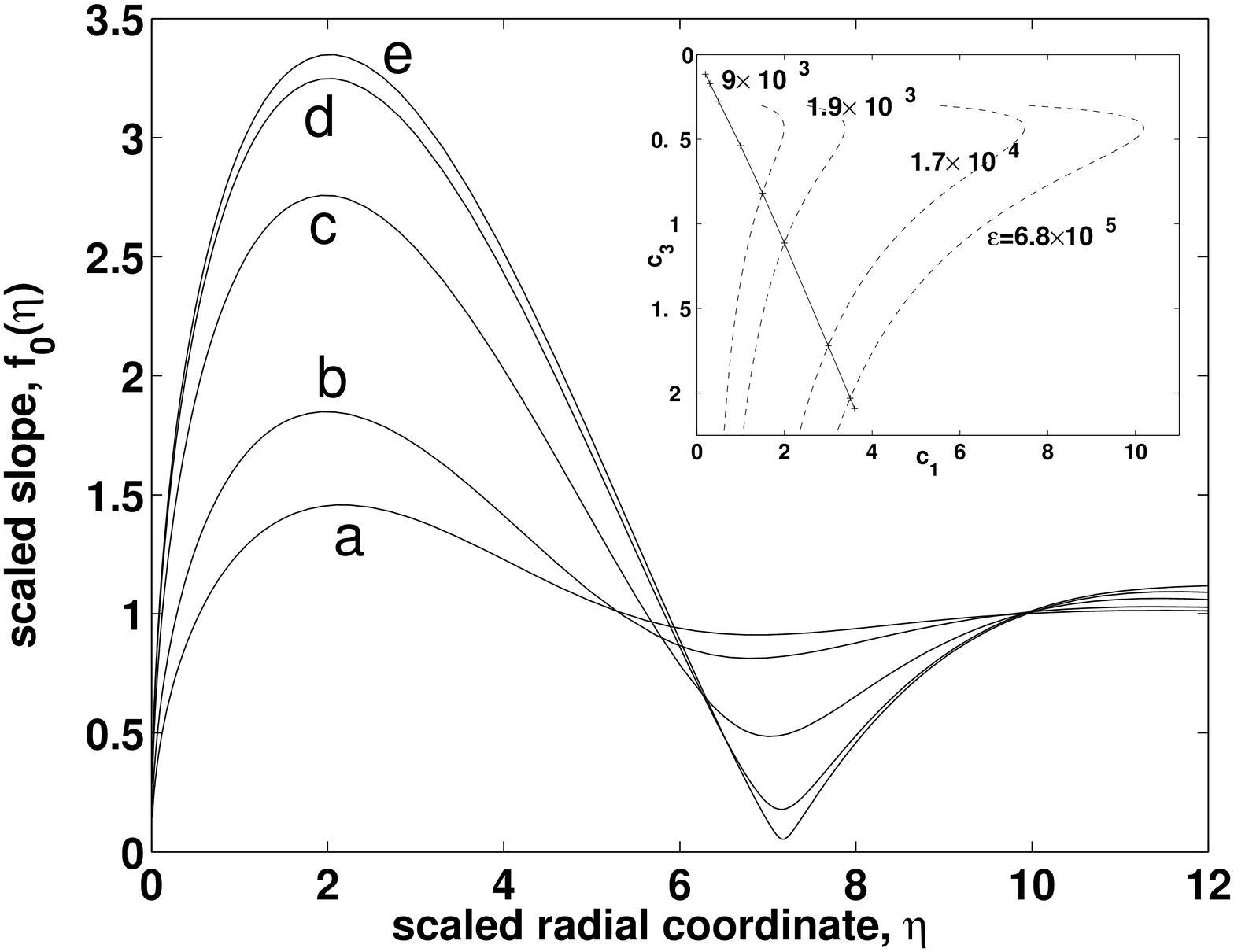}}
\caption{\label{fig2}\mbox{Numerical solutions of
Eq.~(\ref{f0-eqn2}) with the} \mbox{boundary conditions $f_0(0)=0$
and $\ f_0(\infty)=1$. Curves a-e} \mbox{are parametrized by
$(c_1,c_3)=(1.5,-.8183548)$,$\,(2,-1.113031)$},
\mbox{$\,(3,-1.72107502)$,
$\,(3.5,-2.0302102)$,$\,(3.6,-2.09232155)$} and correspond to
$\epsilon=9.2\times 10^{-3}$,$\,1.9\times 10^{-3}$, $\,1.7\times
10^{-4}$, $\,6.8\times 10^{-5}$,$\,5.7\times 10^{-5}$. Inset: The
dashed curves are described by Eq.~(\ref{c1c3-curve}) for a
conical initial shape and different $\epsilon$, while the solid
curve shows $c_3$ as a function of $c_1$ from the numerical
solutions of Eq.~(\ref{f0-eqn2}).}
\end{figure}

We next examine solutions of Eq.\ (\ref{f0-eqn1}) along with the prescribed boundary conditions.
First, this equation is integrated once via
matching ${\cal F}(\eta,t)$ with the outer solution
(\ref{F0-eqn}), i.e., taking $\eta\gg 1$ and $r\to w^+$
simultaneously. We find $a_0(t) = 3Bt w^{-4}-H^{\prime}(w)$,
which by Eq.\ (\ref{similarity-soln}) determines
the explicit time dependence of the surface slope.
Because at this point we have imposed no restrictions other than
axisymmetry on the initial shape, we have in fact obtained a
universal equation for $f_0(\eta)$, i.e.,
\begin{equation}
(f_0^2)^{\prime\prime\prime} = f_0-1~,
\label{f0-eqn2}
\end{equation}which is to be solved with $f_0(0) = 0$
and $f_0(\eta\rightarrow\infty)=1$.
Near the origin,$\,f_0(\eta)$ has the behavior
\begin{equation}
f_0(\eta)\sim c_1 \eta^{1/2}+c_3 \eta^{3/2} +c_5 \eta^{5/2}+c_6
\eta^{3}+...,\label{asympt-behav}
\end{equation}
where all coefficients $c_n$ with $n\ge 5$ are known in terms of
$c_1$ and $c_3$. Equation (\ref{f0-eqn2})
has a growing mode for $\eta \gg 1$, which must be suppressed in order
to satisfy the condition at $\infty$; thus, $c_3$ is found numerically in terms of $c_1$.
We solve Eq.\ (\ref{f0-eqn2}) numerically and so obtain a
family of similarity solutions $f_0(\eta)$ for different values of $c_1$~\cite{comment};
see Fig.~2. We next show how
the solution $f_0(\eta)$ depends on $\epsilon$, which requires imposing conditions such as (\ref{conds}).

The substitution of Eq.\ (\ref{similarity-soln}) into Eq.\
(\ref{conds}) and use of the relations $(f_0^2)'_{\eta=0}=c_1^2$
and $(f_0^2)''_{\eta}=4c_1c_3$ from Eq.\ (\ref{asympt-behav})
yield two parametric equations for $c_1$ and $c_3$. In the case
with a conical initial shape, discussed at length below,
continuity of the variable $\mu$ implies~\cite{margetisetal}
\begin{equation}
(c_1c_3)\epsilon^{1/3}=-\textstyle{\frac{3}{4^{5/3}}}
[c_3^{-2}(c_3^2-\frac{1}{16})^2(c_3^2+\frac{3}{16})^{-1}]^{1/3}.\label{c1c3-curve}
\end{equation}
The intersection of the curve (\ref{c1c3-curve}) with the set of points ($c_1,c_3$)
that result from numerically solving Eq.~(\ref{f0-eqn2}) is shown in the inset of
Fig.~2, and determines a value of $\epsilon$ for each of the solution curves of
the main part of the figure. Thus, we have determined a family of $\epsilon$-dependent
similarity solutions $f_0(\eta;\epsilon)$.

There is one more scaling law that comes from the analysis. Each
of the curves $f_0(\eta)$ in Fig.~2 has a well-defined absolute
maximum. Using (\ref{asympt-behav}) each maximum may be estimated
to be $O(c_1^{3/2}|c_3|^{-1/2})$ and to occur at
$\eta_{max}=O(c_1/|c_3|)$, which is independent of $\epsilon$ to
leading order. Thus, according to Eq.\ (\ref{c1c3-curve}),$\,c_1$
and $c_3$ are $O(\epsilon^{-1/6})$ and so the maximum slope is predicted
to be $O(\epsilon^{-1/6})$.

We now compare the predictions from this continuum approach based
on Eq.\ (\ref{governing-eqn}) with the kinetic simulations for the
DL case reported by Israeli and Kandel~\cite{israelikandel99} for
a conical initial shape. In their simulations these authors vary a
parameter, $\,g=({\rm const.})\cdot g_3$, holding $g_1$ fixed,
which in our analysis is equivalent to changing $\epsilon$. Their
simulations furnished a $g$-dependent family of solutions (see
their Figs.~4(b) and 6), which correspond to our
$\epsilon$-dependent curves $f_0(\eta;\epsilon)$. Israeli and Kandel also
derived a complicated, $g$-dependent differential equation in the
scaling variable $x=r/(Bt)^{1/4}$ (not to be confused with
the cartesian coordinate), which, in the limit of small
$g$, effectively reduces to our (\ref{f0-eqn2}). However, on the
basis of their equation they found multiple solutions whereas we provide a unique solution
$f_0(\eta;\epsilon)$ for each $\epsilon$.

\begin{figure}
\centerline{\epsfxsize0.35\textwidth\epsfbox{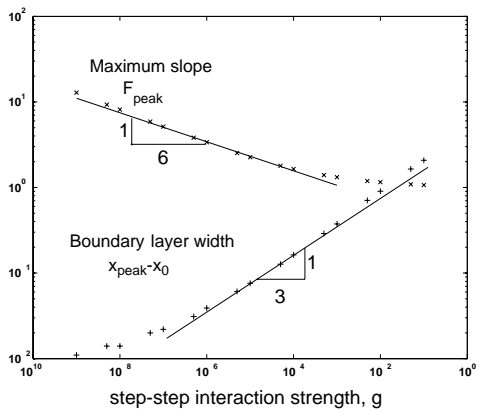}}
\caption{\label{fig3}Log-log plot of the  boundary-layer thickness
$\delta(t; \epsilon)$ and the maximum of step density $F_{\rm
peak}$ as functions of $\epsilon$. The crosses represent the
results of kinetic simulations given to us by Israeli and
Kandel~\cite{israelikandel99} for the DL case. Here, $\,\delta(t;
\epsilon)$ is estimated as the distance $x_{\rm peak}-x_0$ between
the facet edge,$\,x_0=w(Bt)^{-1/4}$, where $F=0$, and the position $x_{\rm peak}$
of the maximum of $F$. The straight lines correspond
to the $\epsilon^{1/3}$  and $\epsilon^{-1/6}$ scaling laws
predicted according to Eq.\ (\ref{f0-eqn1}).}
\end{figure}

Next, we consider scaling behavior with $\epsilon$. First, we
examine the scaling of the boundary layer near the facet edge. We
define the boundary-layer thickness as the distance from the facet
edge,$\,x_0$, to the position of the peak,$\,x_{\rm peak}$, of the step density,$\,F_{\rm peak}$. In Fig.~3 we show
the results of kinetic simulations (symbols) for $x_{\rm peak}-x_0$ vs. $g$ and compare
with our $\epsilon^{1/3}$ scaling prediction (solid line).
Second, in Fig.~3
we examine how $F_{\rm peak}$ varies with $g$, for which
the results of kinetic simulations (symbols) are compared
with the $\epsilon^{-1/6}$ scaling prediction (solid line).
In both cases the agreement is very good.
With regard to the deviations in the boundary-layer width
for $g<10^{-6}$,
as $\epsilon$ decreases in the simulations $x_{\rm peak}$
approaches the facet edge so that the boundary-layer width is
relatively small on the scale of the step spacing and is
consequently poorly defined; its evaluation in discrete
simulations thus becomes prone to errors.

As shown above, qualitative predictions, such as the form of the
multiple solution curves, and quantitative predictions, such as
the $\epsilon^{1/3}$ scaling of the boundary-layer width and the
$\epsilon^{-1/6}$ scaling of the maximum of the slope,
can be deduced from a continuum approach based on Eq.\
(\ref{governing-eqn}) with the use of free-boundary
and boundary-layer theories~\cite{murtynote}.
Further, simple analytical arguments show
that, for any admissible initial slope $F(r,0)=\kappa r^{\nu}$,
for a wide range of $\nu$ including $-4<\nu \le 1$, the facet radius is
$w=O(t^{1/(\nu+4)})$ at sufficiently long times.
In addition, we expect that, for a class
of non-axisymmetric initial shapes, the near-facet boundary layer width
still retains the $O(\epsilon^{1/3})$ scaling for the isotropic
surface free energy of Eq.\ (\ref{G-eqn}).

The continuum approach and the free-boundary viewpoint
capture the essential physics of crystal surface evolution below
$T_R$. This development should give further impetus to continuum
approaches to morphological evolution even at the nanoscale for
structures far below $T_R$.

\acknowledgments {This research was supported by the Harvard NSEC.
We thank M.~Z.\ Bazant and R.~R. Rosales
for useful
discussions, and also thank N. Israeli and D. Kandel for valuable
feedback and for sharing with us details of their simulation
results.}





\begin{thebibliography}{[30]}
\bibitem{shenoyfreund02}Our boundary conditions appear to be consistent with the global variational approach taken
by~\cite{shenoyfreund}.
\bibitem{comment}We solve Eq.\ (\ref{f0-eqn2}) numerically by integrating from $\eta_0\ll 1$
towards $\eta^*\gg 1$ for fixed $c_1$ to satisfy $f_0(\eta^*)= 1$.
This procedure yields a family of similarity solutions
$f_0(\eta)$ corresponding to a curve $c_3(c_1)$ where
$c_1>0$ and $c_3<0$.
\end{thebibliography}
\end{document}